# OPTIMIZING MAGNETORESISTIVE SENSOR SIGNAL-TO-NOISE VIA PINNING FIELD TUNING

J. Moulin [1], A. Doll [1], E. Paul [1], M. Pannetier-Lecoeur [1], C. Fermon [1], N. Sergeeva-Chollet [2], A. Solignac [1]

*1. SPEC, CEA, CNRS, Université Paris-Saclay, CEA Saclay 91191 Gif-sur-Yvette Cedex, France*
*2. CEA LIST, 91191 Gif-sur-Yvette, France*

The presence of magnetic noise in magnetoresistive-based magnetic sensors degrades their detection limit at low frequencies. In this paper, different ways of stabilizing the magnetic sensing layer to suppress magnetic noise are investigated by applying a pinning field, either by an external field, internally in the stack or by shape anisotropy. We show that these three methods are equivalent, could be combined and that there is a competition between noise suppression and sensitivity reduction, which results in an optimum total pinning field, for which the detection limit of the sensor is improved up to a factor of ten.

* Electronic mail: julien.moulin@cea.fr

Thanks to a typical detection level of a few nanoteslas and a large frequency range, magnetoresistive sensors are widely used for weak magnetic fields measurements[1], for example in automotive[2] or biological systems[3,4]. Nevertheless, their limit of detection is often constrained by the presence of Random Telegraphic Noise (RTN) or 1/f low frequency magnetic noise due to domain fluctuations in the magnetic layers[5–7]. Giant MagnetoResistive (GMR) sensors are composed of two magnetic layers separated by a metallic spacer. The reference layer possesses a fixed magnetization while the free layer magnetization rotates as a function of the external magnetic field and induces a resistance variation of the structure due to spin-dependent charge transport[1,8]. In order to obtain a linear resistance variation to an external field, a weak anisotropy or pinning in the free layer needs to be created at 90° from the reference layer magnetization[9]. Furthermore, strong pinning of the free layer has been shown to suppress magnetic noise by magnetization stabilization[10–14]. In this paper, we study the impact of free layer pinning on the sensitivity and the noise behaviour of the GMR in the regime of low pinning fields. In particular, we highlight the existence of an optimum pinning field which suppresses the magnetic noise with a limited sensitivity reduction. This optimum pinning field leads to a detectivity improvement by up to a factor ten. The pinning is applied in three different ways: by an external field (method 1), by an internal coupling inside the stack (method 2) and by the sensor's shape anisotropy (method 3).

In order to experimentally investigate the pinning of the free layer, we use spin valve sensors deposited by sputtering on thermally oxidized (500 nm $SiO_2$) silicon wafers. Two types of stack structures are used to test the aforementioned three ways of free layer pinning. A typical spin valve sensor stack (stack 1) allows studying the effect of the external field and shape anisotropy pinning. A batch of pinned spin valve sensors (stack 2) is deposited to investigate the intrinsic pinning. The pinning strength is applied in the direction of free layer and therefore at 90° from the direction of the reference layer. This reference layer direction is also the GMR sensitivity axis, as depicted in the insert of FIG. 1. For each stack, the reference layer is composed of $Co_{90}Fe_{10}$ in a synthetic antiferromagnet (SAF) structure that is exchange-biased by an adjacent antiferromagnet[15,16]. Stack 1 has the following structure: Ta (3)/$Ni_{89}Fe_{19}$ (3.5)/$Co_{90}Fe_{10}$ (1.5)/Cu (2.3)/$Co_{90}Fe_{10}$ (2.1)/Ru (0.85)/$Co_{90}Fe_{10}$ (2)/$Pt_{38}Mn_{62}$ (18)/Ta (3) (thicknesses in nanometers). This stack has been used for methods 1 and 3.

The generic layer sequence of stack 2 is Ru (1)/$Pt_{38}Mn_{62}$ (18)/$Co_{90}Fe_{10}$ (2)/Ru (0.85)/$Co_{90}Fe_{10}$ (2.1)/Cu (2.3)/$Co_{90}Fe_{10}$ (1)/$Ni_{89}Fe_{19}$ (4.5)/Ru ($t_{Ru}$)/$Co_{90}Fe_{10}$ (2)/$Pt_{38}Mn_{62}$ (11)/Ta (3)/Ru (3). Each stack differs from the others by the thickness of the ruthenium spacer ($t_{Ru}$) from 1.7 nm to 2.8 nm, controlling the RKKY coupling intensity[17,18] between the free layer and the pinning layer. The stacks are annealed for 1 hour at 300° C under 1 T magnetic field to orient the reference layer. For stack 2, a second annealing step for 10 min at 80 mT and 300° C orients the free layer pinning at 90° from the reference layer (see insert FIG. 1). The field applied during the second annealing is chosen to be strong enough to orient the free layer but low enough to have no impact on the reference layer orientation. As the antiferromagnet in the free and the reference layer is the same, the temperature of the two annealings needed to be the same to orient the exchange bias[19,20]. The pinning strength of stack 2 is measured by the sensor response to an applied effective magnetic field along the free layer pinning direction and ranges between 1.1 and 9.6 mT. Stack 2 has been used for method 2.

The spin valves are patterned by optical lithography. All our sensors are yoke-shaped devices[21,22], etched by Ar ion milling. This geometry provides reduced low-frequency noise levels by stabilizing the magnetic domain structure inside the main arm of the yoke[23]. In this work, sensors are 1 µm to 20 µm wide, with an aspect ratio of 50:1. The yokes are connected in a current-in-plane (CIP) configuration by Ta (5)/Cu (150)/Ta (5) contacts and passivated by a protective 150 nm thick $Al_2O_3$ layer. The sensor resistances at zero field are around 1 kΩ for stack 1 devices and 800 Ω for stack 2 devices. The measurements are performed with a feeding current between 1 mA and 3 mA. The Oersted field created by this current in the GMR is calculated to be in the order of tens of µT and possess components along the out of plane direction and along the GMR sensitivity axis but not along the free layer pinning axis. We are thus not considering further the Oersted field in this paper.

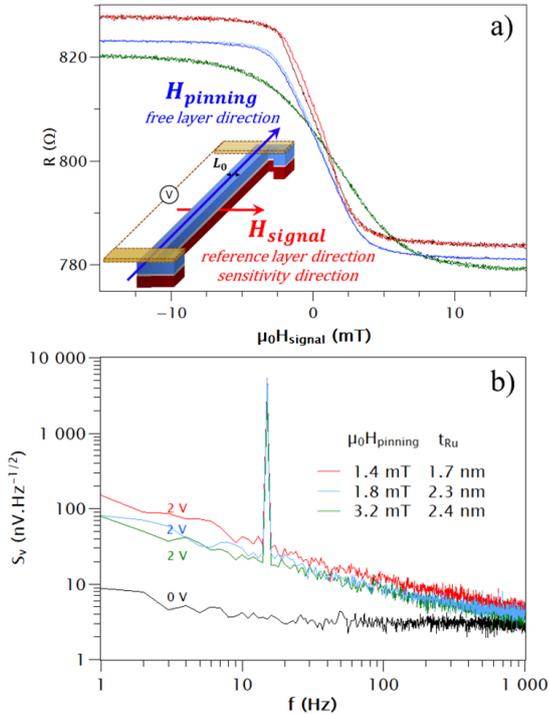

FIG. 1(a) Resistance response to an applied magnetic signal along the sensitive axis for three stack 2 sensors obtained by using different thickness of Ru in the free layer leading to different pinning strengths (geometry 5x250 µm²), legend according to panel b. The inserted sketch illustrates the yoke shaped sensor configuration. Two contacts (beige) are used for resistance measurements. (b) Associated noise spectral density versus frequency. The reference peak at 15 Hz corresponds to a 850 nT$_{rms}$ magnetic signal applied along the sensitivity axis. Sensors are fed with 2 V voltage bias. The black reference curve is the spectral noise density measured on the 2.1 mT-pinned GMR fed with a 0 V voltage bias. The noise floor is the addition of the GMR thermal noise and the noise coming from the amplifier, typically 1 nV/√Hz.

We have then studied the impact of free layer pinning on the GMR sensitivity and noise. We will first focus on method 2 where the pinning is created by an internal pinned layer at 90°. FIG. 1 shows typical resistance and noise GMR responses which allow to extract two important sensor parameters, namely the GMR sensitivity in V/V/T and the noise level in V/√Hz. The sensitivity is defined as the slope of the resistance versus field curve (FIG. 1(a)) around zero field. For the noise measurements[24], we bias GMR sensors using a battery through a balanced Wheatstone bridge. The bridge output is amplified by an INA103 low-pass amplifier before a second step of amplification and band-pass filtering. The whole setup is shielded in a mu-metal magnetic room. An acquisition card acquires the temporal signal and a Fast Fourier transform (FFT) is used to measure the noise spectral density. The measured noise spectral density (FIG. 1(b)) has several components[24]: low frequency 1/f noise, Lorentzian RTN and thermal white noise. Low frequency noise has an electric and a magnetic contribution, where the latter is generally attributed to magnetic domain fluctuations. An AC field signal created by a coil (850 nT$_{rms}$, 15 Hz) and applied along the sensitivity axis serves as calibration reference. It is a complementary way to determine the sensor sensitivity.

The three curves shown in FIG. 1 originate from stack 2 sensors with pinning strength ranging from 1.4 mT to 3.2 mT (see legend). The gradual reduction of the GMR sensitivity with increased pinning is readily identified by the reduction of the slope around the center in FIG. 1(a). In addition, it can be seen that the magnetic hysteresis, which originates from magnetocrystalline anisotropy in the free layer, is suppressed when increasing the pinning strength. For the noise data in FIG. 1(b), a significant reduction in low-frequency noise is observed when increasing the pinning field from 1.4 mT to 1.8 mT (red to blue). However, a further increase of the pinning field from 1.8 mT to 3.2 mT (green) does not induce a significant change in the noise spectrum. From these three curves, one can therefore already deduce that the signal-to-noise ratio of the sensor has a non-linear dependence on the pinning field.

For a more complete picture, the influence of the pinning field on the sensitivity and on the noise level at 30 Hz is shown in FIG. 2(a) for ordinary spin-valve sensors (stack 1). The pinning is in this case (method 1) applied by an external field created by a coil. The coil possesses a yoke shape $Ni_{89}Fe_{19}$ core to concentrate the field on the GMR. Note that we observe a symmetric and non-hysteretic behaviour for both positive and negative pinning field values. This indicates that the Oersted field created by the GMR current supply is negligible in our data. As expected by the Stoner-Wohlfarth macrospin model[10,25] (solid blue line FIG. 2(a)), the sensitivity decreases continuously as an inverse function of the applied pinning field. For the noise, we observe a sharp reduction for a pinning field of 2 mT. We attribute this behaviour to magnetic noise (1/f and RTN) suppression down to the electric noise level due to magnetic domain stabilization in the free layer by the pinning field.

FIG. 2(b) depicts the field equivalent noise versus the pinning field (blue). The performance of the sensor is determined by the field equivalent noise level (also called detectivity) in T/√Hz and is calculated as the ratio between the noise spectral density and the sensitivity. It corresponds also to the field at which the signal to noise ratio is equal to one for a one-second acquisition. As is readily seen, there is an optimum pinning field value determined by the competition between the magnetic noise reduction and the sensitivity loss as the pinning field increases. Typical detection limit enhancement factors are between 3 and 10, depending on the frequency. As an example, here in stack 1 for method 1, at 30 Hz, the detectivity is driven by a 2.75 mT effective pinning field from about 30 nT/√Hz to 5 nT/√Hz, with a strong repeatability improvement due to RTN suppression.

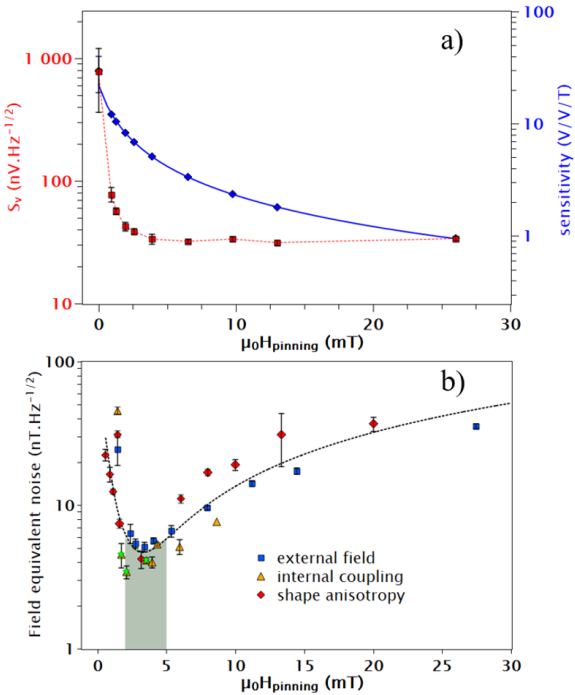

FIG. 2(a) Noise effective value (at 30 Hz) and field sensitivity as a function of magnetic external pinning field (stack 1 – method 1), the blue line is the fitted tendency according to the macrospin Stoner-Wohlfarth model[10,25]. (b) Field equivalent noise at 30 Hz as a function of the effective pinning field for the three pinning strategies: shape, external and internal field. The sensor volume is normalized to a 5 µm width GMR sensor. Green stars highlight the stack 2 devices described in Fig. 1. The data point at lowest pinning of 1.15 mT for stack 2 was realized by compensation of the internal coupling using method 1. The hatched area shows the typical field range where the optimum is located. Dashed lines are guides to the eye.

The detectivity optimization by a pinning field is only possible when magnetic 1/f and RTN noise are present. An optimum can then be found for the entire magnetic noise frequency range. We relate the presence of the magnetic noise to the presence of hysteretic behaviour of the MR sensor response (see FIG. 1(a)) and we observe a direct link between the opening of the hysteresis cycle and the amplitude of the magnetic noise. At low pinning (1.4 mT), the high sensitivity is deteriorated by high low-frequency noise and an open hysteresis cycle. At higher pinning (3.2 mT), the hysteresis and the magnetic noise are suppressed but the sensitivity is visibly reduced. At the optimal effective total pinning (2.25 mT for stack 2), the best detectivity is reached thanks to magnetic noise suppression and a small sensitivity decrease. In this case, an enhancement factor of 12.5 is observed with respect to a stack 2 sensor with zero net pinning.

Besides the results with method 1 discussed above, FIG. 2(b) also shows profiles for pinning methods 2 (yellow) and 3 (red). As a reminder, for method 1, we apply a controlled external magnetic field in the direction perpendicular to the sensitivity axis of the yoke. For method 2 the pinning field is applied inside the stack through RKKY coupling. For method 3, the pinning strength is applied by shape anisotropy and controlled by sensor width variation from 1 µm to 20 µm. Accordingly, the presence of the optimum pinning field is observed for the three different pinning strategies. The detectivity versus the effective pinning field follows the same trend with an improved detectivity for an optimum pinning field value, typically between 2 and 5 mT, depending on the stack structure.

It is important to mention that the shape anisotropy is present for each strategy and has to be taken into account in the effective pinning estimation. Importantly, it is possible to combine the pinning strategies and to adapt it to the application targeted. When a small permanent magnet could be pasted close to the GMR, in terms of size and sample stray field perturbation[26], the optimal pinning could be applied with an external field. It is also possible to control the direction of the strong perpendicular field needed in certain applications, such as magnetic resonance detection[27] or the detection of cells using magnetic nanoparticles[28], so that the residual in-plane field establishes an optimum pinning. When it is not possible to apply an external field, shape anisotropy or internal stack pinning could be used as for instance for sensors in automotive applications.

Despite the fact that all three methods contribute equally to the effective pinning, it is important to highlight that the shape anisotropy pinning creates an axial pinning whereas the external field and the internal stack pinning create a directional pinning. Hence, sensors pinned by shape anisotropy are more sensitive to magnetic history due to the presence of two stable magnetic positions contrary to the two other methods.

Targeting practical uses, we will now describe how to design the sensor for a given stack and a chosen pinning method. As the effective field is just the linear sum of the various fields applied on the free layer, the principle is simply to apply an effective total field equal to the optimal pinning field. For method 1, the best field intensity in terms of detectivity is 2.75 mT and corresponds to the 1.15 mT shape anisotropy field for a 5 µm sensor added to a 1.6 mT external field which then needs to be applied. For method 2, it is essential to determine the proper thickness of Ru in order to obtain the intern field value of 1.1 mT, which corresponds to the 2.25 mT optimal pinning for this second stack minus the 1.15 mT shape anisotropy field for a 5 µm sensor. Fig. 3(a) gives the RKKY coupling strength using the theoretical model from Bruno et al[17] between two $Co_{90}Fe_{10}$ adjacent layers as a function of the Ru spacer thickness. Three points can be chosen: 1.8 nm, 2.2 nm and 2.5 nm. As the MR ratio is increasing when the thickness of Ru is decreasing, it is preferential to choose 1.8 nm. Experimentally, no significant difference has been observed between antiparallel and parallel pinning configurations for identical absolute pinning.

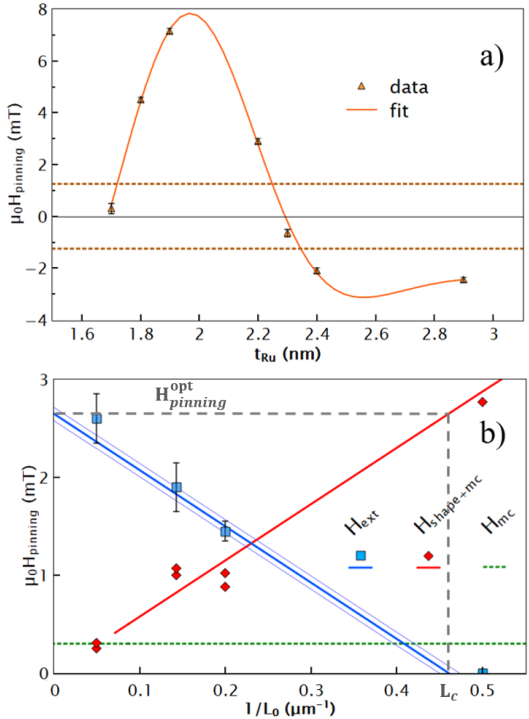

FIG. 3(a) Evolution of the coupling strength in intrinsically pinned devices (stack 2 – geometry 5x250 µm² – method 2) as a function of the ruthenium spacer thickness. Triangles are experimental data. The solid curve is the computed oscillatory "RKKY-like" coupling using the theoretical model from Bruno et al[17]. The intersections between the dashed line, which is the optimal pinning field needed, and the solid line give the Ru thickness to be used. (b) Evolution of the external optimal field as function of the anisotropy field induced by the GMR width $L_0$. Blue squares are experimental optimal pinning measurements, with fitted tendency in solid blue. The red solid line is the computed intrinsic anisotropy field of the sensor using experimental data (red diamonds). The green dashed line is the estimated magnetocrystalline anisotropy value. The grey dashed line corresponds to the graphical solution of Eq. (1).

For method 3, we need to extract the critical width $L_c$ associated to the optimal detection limit $H_{pinning}^{opt}$ using the following equation[29]:

$$L_c = M_s t / H_{pinning}^{opt} \quad (1)$$

$M_s$ and $t$ are the saturation magnetization and the thickness of the free layer, respectively. If we perform that calculation for stack 1, we obtain 2.25 µm as optimal width for a pinning of 2.75 mT. FIG. 3(b) shows the dependence between the pinning field and the width of the sensor. The red line denotes the inverse dependence that follows from a macrospin model and corresponding experimental data points $H_{shape+mc}$ (red diamonds). The optimum external pinning fields $H_{ext}$ for noise reduction (blue squares) superimpose on the blue curve, whose slope is the inverse of the red curve. The intersection point, where no external field is needed also corresponds to the critical width $L_c$ in Eq. (1).

If the stack is strongly modified, the curve given in FIG. 2(b) has to be measured on a sensor to determine $H_{pinning}^{opt}$, the simplest way consisting in using method 1. The pinning strategy, that can be a combination of methods 1, 2 and 3, is then chosen depending on the application targeted.

For further insight into the magnetic noise in our devices, we estimated the size of the magnetic fluctuators responsible for RTN by using a model that relates the amplitude of the RTN voltage fluctuation by a magnetic oscillator of volume $V$ to the magnetoresistance value of the entire GMR device[30]:

$$V \cos(\frac{\pi}{2} - \theta) = \frac{\partial R}{R} \frac{V_{GMR}}{MR_{GMR}} \quad (2)$$

where $\theta$ is the angle between the two magnetization states of the fluctuator. The dependence of the lifetimes $\tau_1$ and $\tau_2$ of the two magnetic states as a function of the magnetic field follows an Arrhenius law[5,31]:

$$\frac{\tau_1}{\tau_2} \propto \exp(\frac{\vec{B}\Delta\vec{m}_{12}}{kT}) \quad (3)$$

Where $\Delta\vec{m}_{12}$ is the magnetization variation vector between the two RTN states. In a crude approximation, its amplitude is equal to $2M_s V \sin^2(\theta/2)$. By tracing the statistics of individual fluctuators as a function of the applied field, the two unknown variables $V$ and $\theta$ in equations (2) and (3) can be determined. As a result, typical sizes of such fluctuators in our devices are experimentally estimated between 400 nm and 1.5 µm. With the estimated activation energy barrier of these fluctuators on the order of kT, these are extremely sensitive to microtesla magnetic fields. Accordingly, magnetic RTN suppression can be related to the field-dependent behaviour of the ensemble of magnetic fluctuators.

In summary, we explored the effect of introducing a well-controlled pinning on the free layer, perpendicular to the sensitivity direction, on magnetoresistive sensing performance, focusing on the magnetic noise behaviour. In particular, we found the presence of an optimal pinning field that does not depend on how the pinning is performed. In this way, the magnetic noise is suppressed before losing sensitivity due to over-pinning of the free layer, which results in an enhanced detection limit up to a factor of 10. The effective pinning field is a linear sum of all pinning fields created by shape, structure and external field sources. For a specific sensing application, we can thus combine either of these methods to "build" the best-adapted sensor.

We acknowledge the following organisms for funding:
- The CEA for the internal funded projects MIMOSA and CALM and the PhD Grant "Phare Amont-Aval".
- The Swiss National Science Foundation for a mobility fellowship (165238 and 177732) to A. Doll.
- ANR funding through grants n°ANR-17-CE19-0021-01 (NeuroTMR) and n°ANR-18-CE42-0001 (CARAMEL).
- This work was supported by the EMPIR JRP 15SIB06 NanoMag through EU and EMPIR participating countries within EURAMET.